\documentclass{article}
\usepackage{spconf,amsmath,graphicx}

\usepackage{multirow}
\usepackage{subfig}
\usepackage{url}
\usepackage{booktabs}

\title{Investigations on Audiovisual Emotion Recognition in Noisy Conditions}
\name{Michael Neumann, Ngoc Thang Vu}
\address{University of Stuttgart, Germany\\
    Institute for Natural Language Processing (IMS)\\
    michael.neumann@ims.uni-stuttgart.de}

\begin{document}
\topmargin=0mm
\maketitle

\begin{abstract}
In this paper we explore audiovisual emotion recognition under noisy acoustic conditions with a focus on speech features. We attempt to answer the following research questions:
(i) How does speech emotion recognition perform on noisy data? and
(ii) To what extend does a multimodal approach improve the accuracy and compensate for potential performance degradation at different noise levels?

We present an analytical investigation on two emotion datasets with superimposed noise at different signal-to-noise ratios, comparing three types of acoustic features. Visual features are incorporated with a hybrid fusion approach: The first neural network layers are separate modality-specific ones, followed by at least one shared layer before the final prediction.
The results show a significant performance decrease when a model trained on clean audio is applied to noisy data  and that the addition of visual features alleviates this effect.
\end{abstract}

\section{Introduction and Related Work}

Automatic recognition of human affect has become a fast-growing research field in recent years, finding its way into a variety of application areas, such as health care, call centers, voice assistants, or automotive applications.
While the research focus has long been on one modality individually (e.g. speech or facial expressions), multimodal emotion recognition -- and multimodal machine learning in general -- has gained increasing attention recently.
For a comprehensive survey on multimodal machine learning the reader can refer to~\cite{baltruvsaitis2019multimodal}.
Regarding emotion recognition, Sebe et al. presented a survey on multimodal approaches~\cite{sebe2005multimodal} and proposed probabilistic graphical models for fusing modalities. In another early work, Busso et al. compared early and late fusion and showed that acoustic and visual features contain complementary information about expressed emotions~\cite{busso2004analysis}.
More recently, deep learning and end-to-end learning gained traction in the field, partially because of the availability of larger amounts of training data~\cite{tzirakis2017end, han2019implicit, ghaleb2017multimodal, wollmer2013lstm, mallol2019performance}.

Another aspect of increasing interest is the performance of systems outside of clean laboratory conditions, which is for example addressed by the 'Emotion Recognition in the Wild Challenge'~\cite{dhall2019emotiw}. The effect of noisy data has been investigated for \textit{speech} in several studies~\cite{schuller2006emotion, schuller2007towards, you2006emotion, zhao2014robust} and speech enhancement methods are one promising direction for better SER quality~\cite{chenchah2016speech, avila2018investigating, zhang2016facing, triantafyllopoulos2019towards}. While we are aware of the different methods to attenuate noise effects in speech data, we focus specifically on a multimodal approach because only few studies addressed the problem of noisy data in audiovisual experiments and we want to investigate the complementary effects of both modalities.
Banda et al. focused on the effect of corrupted videos and showed that a multimodal system retains a reasonably high performance compared to video-only~\cite{banda2011noise}. Lin et al. added noise to both audio and video and presented a semi-coupled Hidden Markow Model to diminish the negative impact of noise~\cite{lin2013probabilistic}.
In contrast,  we focus on noisy \textit{acoustics}, leaving the videos untouched. While most of the above-mentioned studies present results for \textit{matched} train and test data (either clean or with added noise) and for positive signal-to-noise ratios (SNR), we also analyze the performance in the \textit{unmatched} condition (train on clean and test on noisy data), including negative SNRs, as it was done by Triantafyllopoulos et al.~\cite{triantafyllopoulos2019towards} with deep learning based speech enhancement.

The main research questions we attempt to answer are:\\
(i) How does speech emotion recognition perform on noisy data (at different noise intensities)?\\
(ii) To what extend does a multimodal approach improve the accuracy in clean and noisy conditions? \\
Our findings show strong performance degradation when speech-only models are applied to noisy audio, throughout different features, noise types and datasets.
Adding visual features and data augmentation both significantly improve the performance.
The analysis of error patterns in predictions reveals a strong bias towards the class \textit{happy} on noisy test data for audio-only; adding visual features weakens this effect. Further, we found that multimodal fusion helps to distinguish the high-arousal emotion classes \textit{angry} and \textit{happy} better from each other in clean audio conditions. Comparing three acoustic feature sets, one main finding is that a convolutional neural network (CNN) with log Mel filterbanks as input performs most stable under noisy conditions, compared to feed-forward networks on the other feature sets, which exhibit strong biases towards single classes. However, the overall best performance is achieved with eGeMAPS features.


\section{Methods}

\subsection{Acoustic Features}
We examine three types of acoustic features, including handcrafted (eGeMAPS) and model-based representations (DeepSpectrum).
Noisy audio is generated as in~\cite{wand2018investigations}: We use the acoustic simulator presented in~\cite{ferras2016large} to superimpose different types of noises from the Freesound database~\cite{font2013freesound} (babble, music and transportation noise) to the clean audio at \{-10, -5, 0, 5\} dB SNR. A sample rate of 16kHz is used for all data.

The extended Geneva minimalistic acoustic parameter set (\textbf{eGeMAPS})~\cite{eyben2016geneva} is a feature set recommendation for affective computing. It consists of 25 low level descriptors (frequency related, energy related and spectral parameters) to which several functionals are applied. This results in an 88-dimensional utterance-level representation. We use openSMILE~\cite{eyben2013recent} to extract eGeMAPS features.


For the third feature set we use \textbf{DeepSpectrum}~\cite{amiriparian2017snore}, a Python toolkit for acoustic feature extraction based on pre-trained image CNNs. Spectrograms are generated from the acoustic signal and fed into a pre-trained CNN. The activations of a specific layer form the feature vectors. We use the DeepSpectrum default settings, i.e. extract the activations of the 'fc2' layer from AlexNet, resulting in a 4,096-dimensional feature vector for one utterance.\footnote{\url{https://github.com/DeepSpectrum/DeepSpectrum}} 


\subsection{Visual Features}
Visual representations are obtained as follows: Each video frame is converted to gray scale and Contrast-limited adaptive histogram equalization~\cite{pizer1987adaptive} is applied to enhance contrast, followed by face recognition using dlib's~\cite{king2009dlib} frontal face detector. The detected face region is cropped and resized to 100x100 pixels to feed it into the VGG 11-layer model (configuration "A" from~\cite{simonyan2014very}) trained on ImageNet~\cite{russakovsky2015imagenet}. We take the activations of the first fully connected layer as frame-level intermediate features and apply average pooling across all frames of the same utterance to obtain a 4,096-dimensional utterance-level representation.

\begin{figure}[htb]
    \centering
    \includegraphics[trim=110 80 85 185, clip, width=\linewidth]{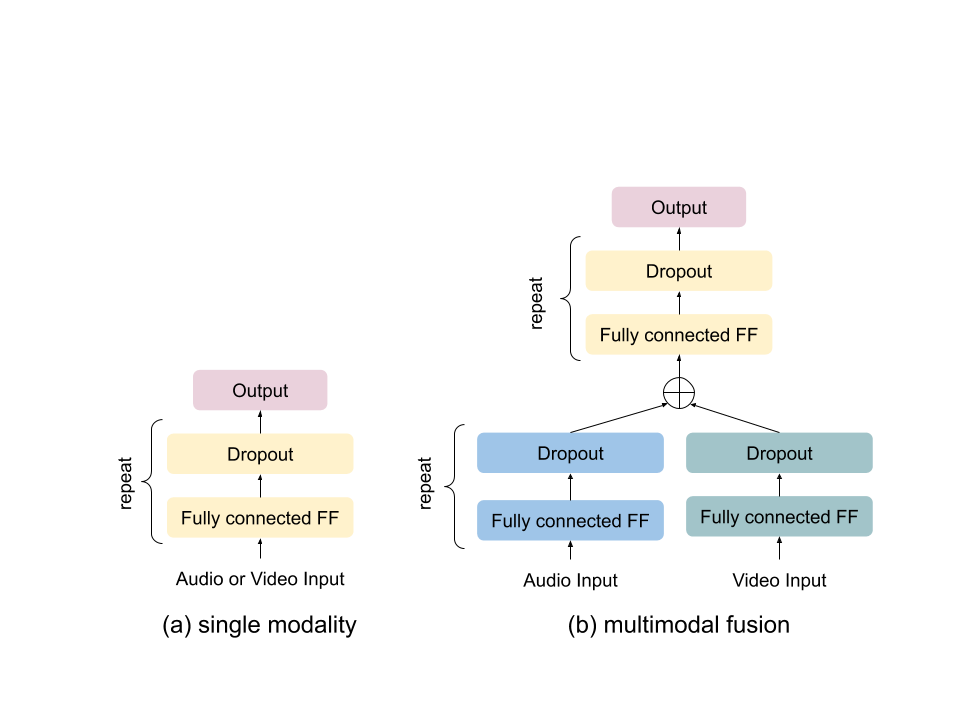}
    \caption{Neural network architectures.}
    \label{fig:networks}
\end{figure}

\subsection{Neural Network Structures}
For both unimodal and multimodal experiments we train fully connected feed-forward (FF) neural networks (except for log Mel filterbanks, for which a CNN is applied). All models are implemented with PyTorch~\cite{NEURIPS2019_9015}. 
The FF networks are composed of a stack of fully connected layers with tanh non-linearity, each followed by dropout regularization, as shown in Fig.~\ref{fig:networks}(a). For filterbanks, which are a time-preserving 3-dimensional matrix, we train a strided CNN composed of two convolutional layers with ReLU activation, each followed by a dropout layer. There is no pooling layer, but pooling is implicitly controlled by tuning the stride size of the convolution. The kernels of the first convolutional layer span the entire input feature dimension (23 filterbanks), i.e. the subsequent layer is a 1-D convolution over time. Because a CNN requires a fixed input size, we set the sample length to 7.5s for MSP-IMPROV and 3s for CREMA-D (based on \textit{mean duration + standard deviation} of each corpus) and apply zero-padding for shorter utterances.

For multimodal fusion we use a hybrid approach, shown in Fig.~\ref{fig:networks}(b): audio and video input is fed into separate sub-networks (at least one layer), whose outputs are concatenated and fed into a joint network. For filterbanks, the sub-network is a CNN (as described above) whose output is flattened and fed into the joint network.

\section{Experimental Setup}
\subsection{Datasets}
The \textbf{MSP-IMPROV} dataset~\cite{busso2017msp} contains English dyadic interactions between actors (six pairs of one female and one male each). Video recordings show the speakers in front of a green screen; audio is provided at 44.1~kHz sample rate.\footnote{MSP-IMPROV audio files were downsampled to 16 kHz} Emotion annotations have been obtained through crowdsourcing (at least five annotations per sample). We use the four discrete emotion classes \textit{angry, happy, neutral, sad}, resulting in a total of 7,798 samples. Due to the imbalanced class distribution\footnote{MSP distribution: 792 angry, 2,644 happy, 3,477 neutral, 885 sad} we use a weighted cross entropy loss function in all experiments. 
The class weights are calculated as $class\_weight = N / (C * N_c)$, where $N$ is the total number of samples, $C$ the number of classes, and $N_c$ the number of samples in class $c$. As there are no default train and test partitions, we apply 6-fold cross validation (leave-one-session-out) to ensure speaker-independent evaluation (10\% of the train set are randomly selected as development set).

The \textbf{CREMA-D} dataset~\cite{cao2014crema} consists of English read speech; 91 speakers read 12 target sentences in six different emotions (\textit{happy, sad, angry, fearful, disgusted}, and \textit{neutral}). Video recordings show the speakers in front of a green screen; audio is provided at 16 kHz sample rate. Annotations have also been obtained through crowdsourcing (at least six per sample). There are individual annotations for audio, video and audiovisual data from which we use the latter. To facilitate comparisons between datasets, we use the same four classes as for MSP-IMPROV, resulting in 4,799 samples. We apply the same weighted loss as described above because the data are also imbalanced.\footnote{CREMA distribution: 1,019 angry, 1,219 happy, 1,972 neutral, 589 sad} For evaluation we split the data by speaker IDs to avoid speaker overlap: speakers 1-63 as train, speakers 64-77 as development, and speakers 78-91 as test set. We have verified that these partitions are balanced regarding age and gender distributions.


\subsection{Hyper-parameters and training}
To establish baselines, we assessed the accuracy on clean audio on the development set (average result across 6 folds for MSP-IMPROV) in a grid search for these hyper-parameters: number of layers, number of neurons per layer, dropout rate (additionally for CNN: number and size of feature maps, stride size).
This was done for each feature type (unimodal and multimodal) and dataset individually.
Table~\ref{tab:params} shows the number and size of layers for each input option.
The CNN hyper-parameters are: 128 feature maps of width 10 and stride 7 for MSP-IMPROV CREMA-D unimodal, and 128 feature maps of width 15 and stride 3 for CREMA-D multimodal.
For MSP-IMPROV, we applied dropout at a rate of 0.5 for all except the DeepSpectrum features, for which we selected 0.7 to prevent overfitting. For CREMA-D, overfitting appears more prevalent, presumably because of the artificial nature of the data. We obtained a dropout rate of 0.7 for all but the unimodal eGeMAPS features, for which 0.5 was applied.
Model training was done with a batch size of 32 for 100 epochs for unimodal and 50 epochs for multimodal input. We ran all experiments (except hyper-parameter tuning) three times and report mean and standard deviation in terms of unweighted average recall (UAR).
\begin{table}[htb]
    \centering
    \resizebox{.85\linewidth}{!}{
    \begin{tabular}{r|c|c||c|c}
         & \multicolumn{2}{|c||}{MSP-IMPROV} & \multicolumn{2}{|c}{CREMA-D} \\
         & \# layers & \# neurons & \# layers & \# neurons \\
         \toprule
        eGeMAPS & 2 & 128 & 3 & 64 \\
        DeepS & 2 & 128 & 2 & 128 \\
        \midrule
        Video (V) & 3 & 256 & 3 & 128 \\
        \midrule
        eGeMAPS+V & 1+1 & 256 & 1+1 & 128 \\
        Filterbanks+V & 2+2 & 256 & 2+1 & 128 \\
        DeepS+V & 1+1 & 128 & 1+1 & 128 \\
    \end{tabular}
    }
    \caption{Hyper-parameters. 'x+y' means: x layers in each sub-network and y layers in the joint network.}
    \label{tab:params}
\end{table}


\section{Results and Analysis}
\label{sec:results}
\subsection{Models trained on clean and applied to noisy audio}
\label{sec:exp1}
\begin{figure*}[ht]
    \centering
    \subfloat[MSP-IMPROV \label{fig:msp_clean}]{\includegraphics[trim=10 10 10 10, clip, width=.5\textwidth]{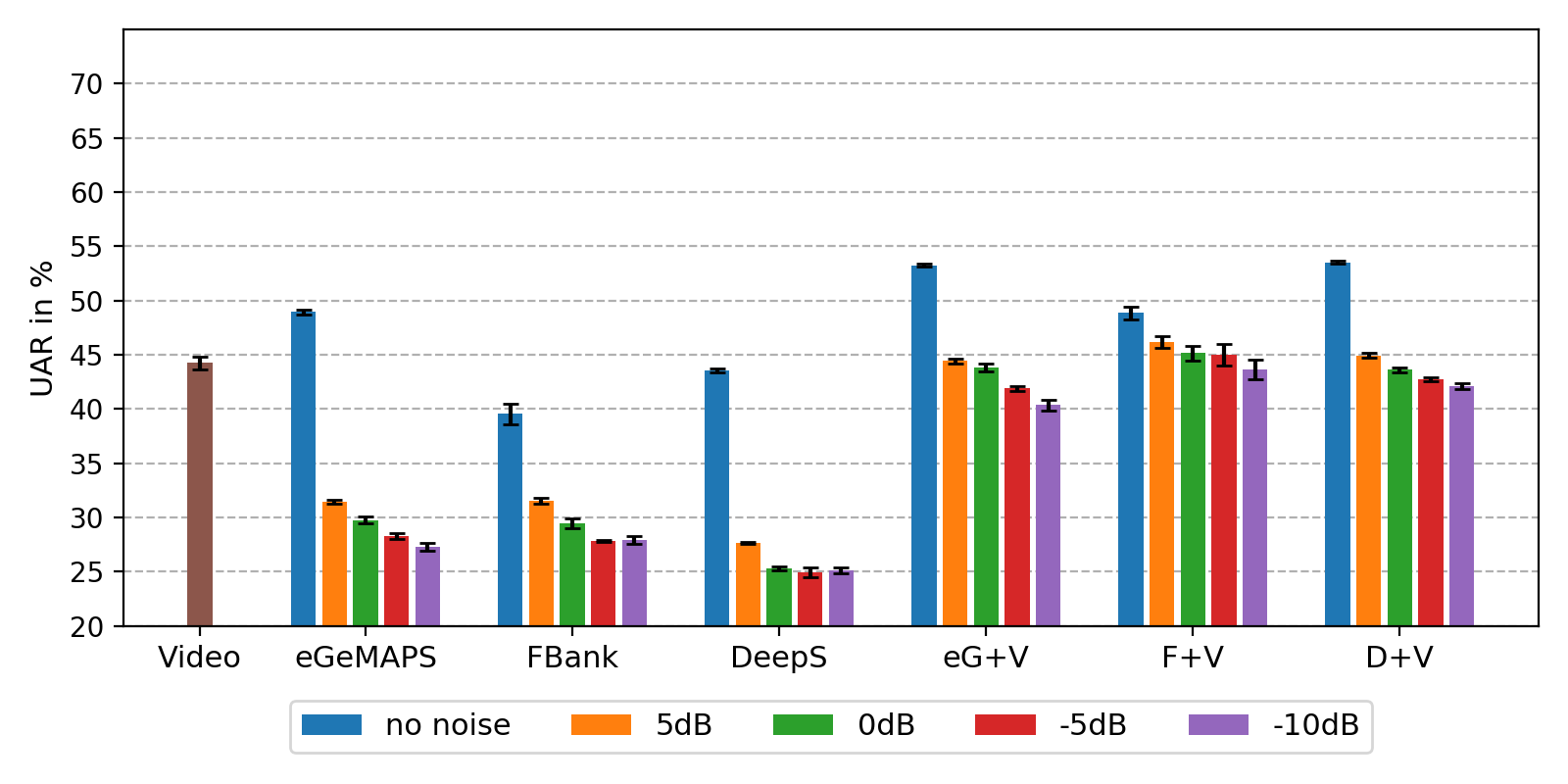}}
    \subfloat[CREMA-D \label{fig:crema_clean}]{\includegraphics[trim=10 10 10 10, clip, width=.5\textwidth]{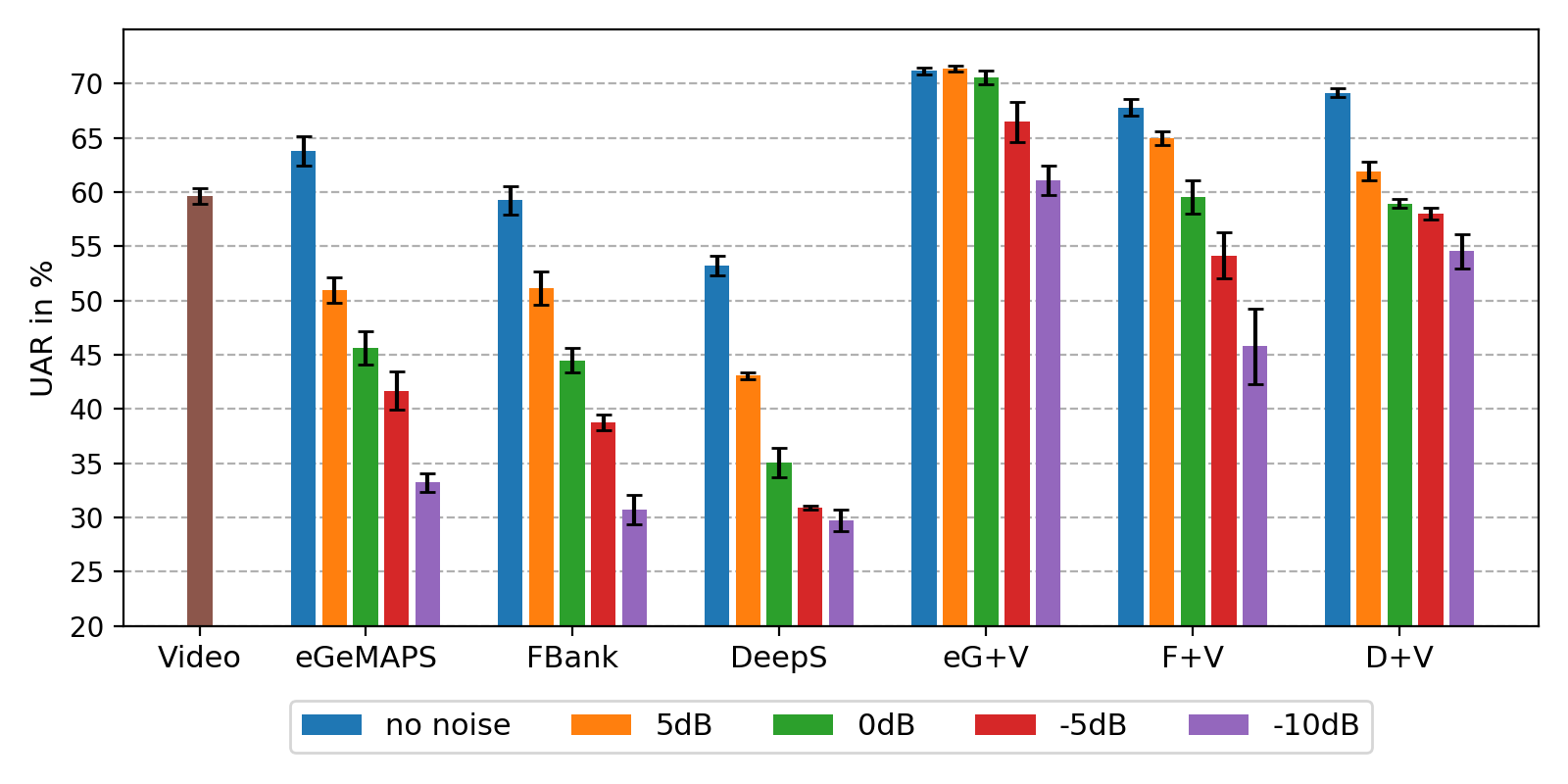}}
    \caption{Results for training on clean audio only and applying the models to noisy audio data (babble noise). \\ FBank (F): filterbanks, DeepS (D): DeepSpectrum, eG: eGeMAPS, V: Video.}
    \label{fig:train_on_clean}
\end{figure*}

\begin{figure}
    \centering
    \includegraphics[trim=10 10 10 10, clip, width=\linewidth]{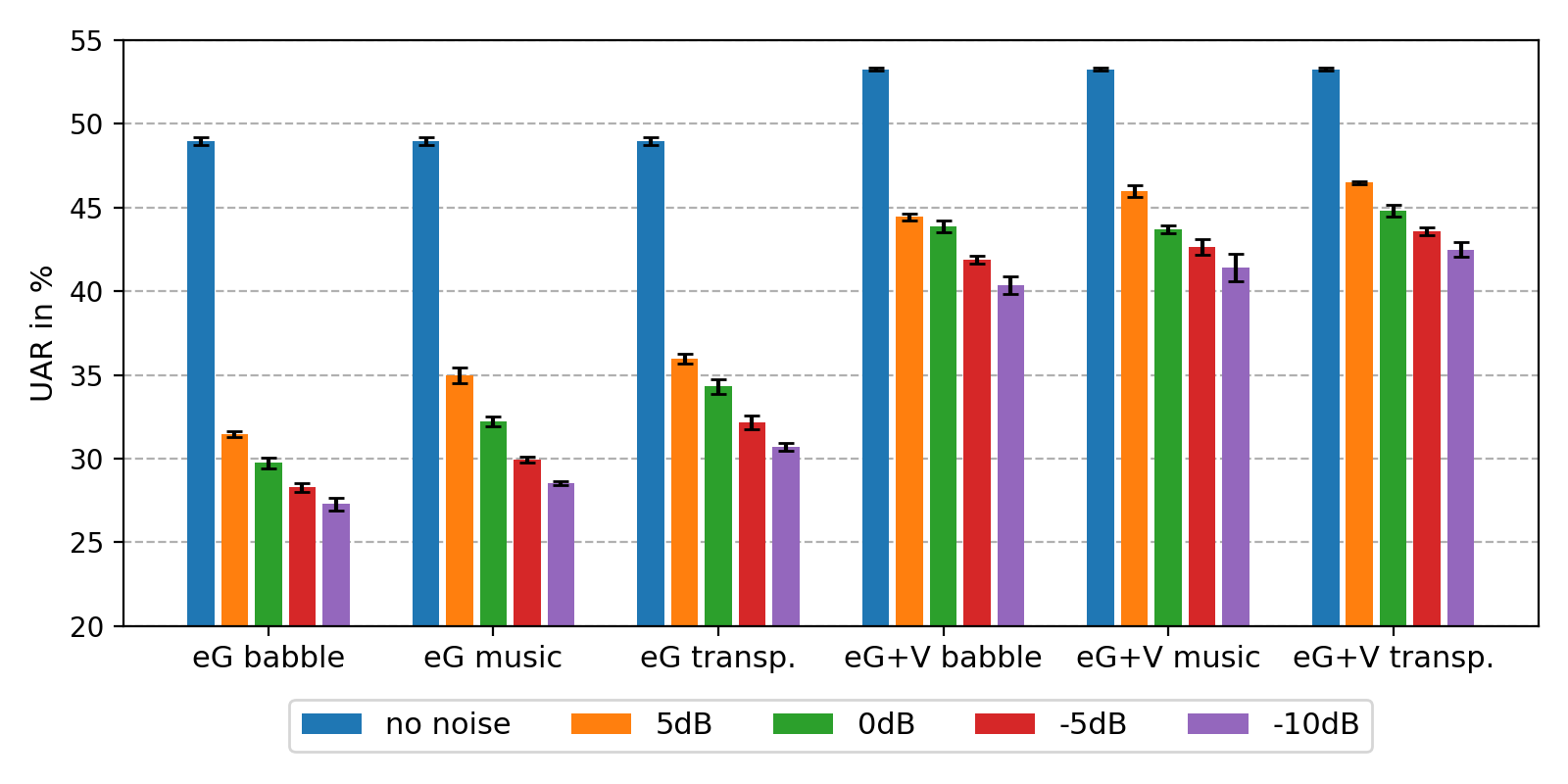}
    \caption{Results on MSP-IMPROV for training with clean audio only and applying the models to noisy audio data (eGeMAPS features, all noise types).}
    \label{fig:msp_clean_all_noises}
\end{figure}

In the first experiment we apply trained models to data with different noise levels and compare the results with the clean reference data.
The results are shown in Fig.~\ref{fig:train_on_clean}.
Comparing different noise types, we observed the same tendencies across features and datasets. Overall, the recognition performance is slightly higher for transportation noise and lowest for babble noise. Fig.~\ref{fig:msp_clean_all_noises} shows results for all three noise types on the MSP-IMPROV data with eGeMAPS features. The remainder of this paper presents results for babble noise only, due to space limitations.

In general, we observed a large decline in performance from clean to noisy audio on both datasets, with decreasing UAR for higher noise levels.
Adding visual features consistently improves the performance by a large margin.

For \textbf{MSP-IMPROV} (Fig.~\ref{fig:msp_clean}), the best audio-only result is 48.59\%~$\pm$0.23\% (clean audio, eGeMAPS), the best multimodal result is 53.50\%~$\pm$0.14\% (clean audio, DeepSpectrum). The video-only UAR is 44.24\%~$\pm$0.59\%.

\begin{figure*}[ht]
    \centering
    \subfloat[eGeMAPS: clean audio\label{fig:gemaps_clean}]{\includegraphics[trim=50 20 85 20, clip, width=.245\textwidth]{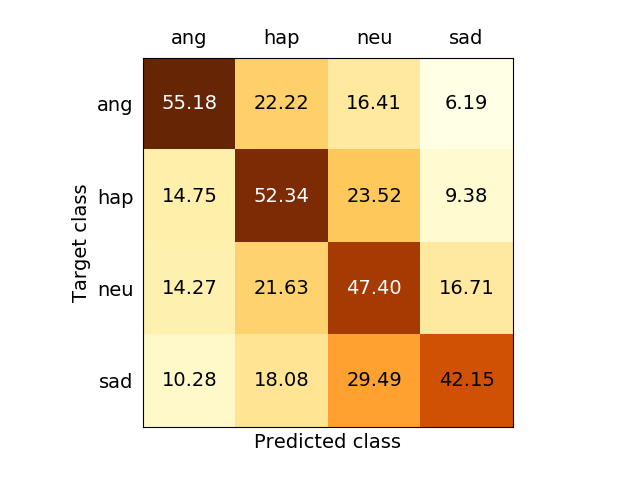}}
    \subfloat[eGeMAPS: 0dB SNR\label{fig:gemaps_snr0}]{\includegraphics[trim=70 20 85 20, clip, width=.23\textwidth]{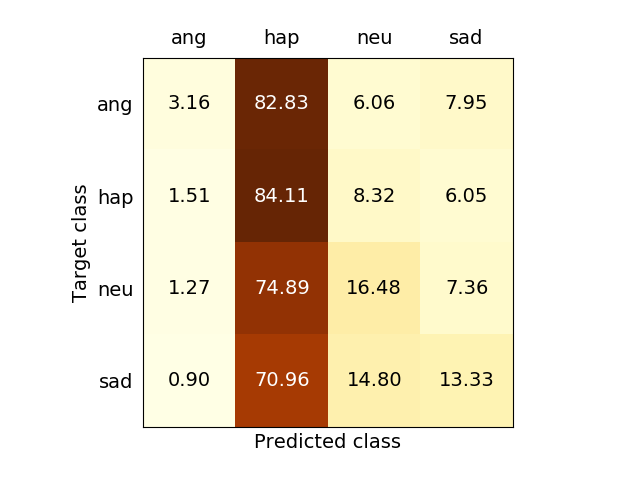}}
    \hfill
    \subfloat[eGeMAPS+Video: clean audio\label{fig:gemaps_clean_multi}]{\includegraphics[trim=50 20 85 20, clip, width=.245\textwidth]{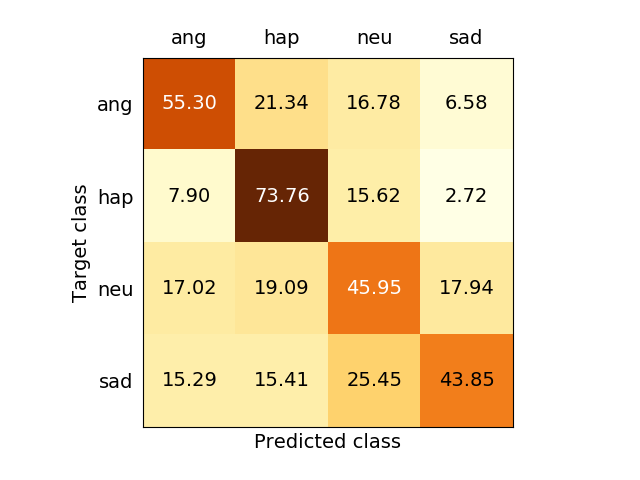}}
    \subfloat[eGeMAPS+Video: 0dB SNR\label{fig:gemaps_snr0_multi}]{\includegraphics[trim=70 20 85 20, clip, width=.23\textwidth]{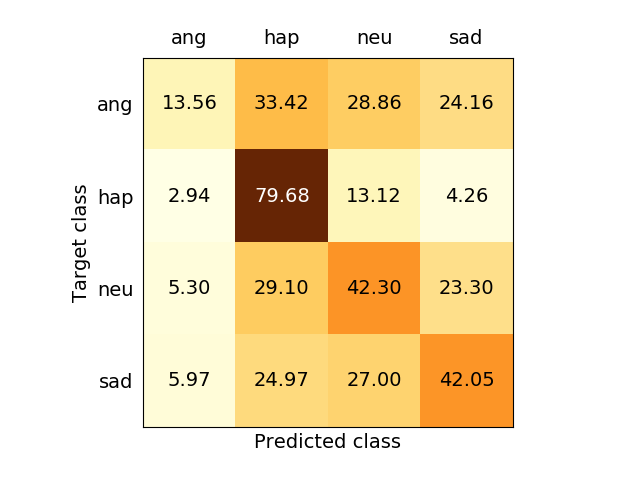}} \\

    \subfloat[FBank: clean audio\label{fig:fbank_clean}]{\includegraphics[trim=50 20 85 20, clip, width=.245\textwidth]{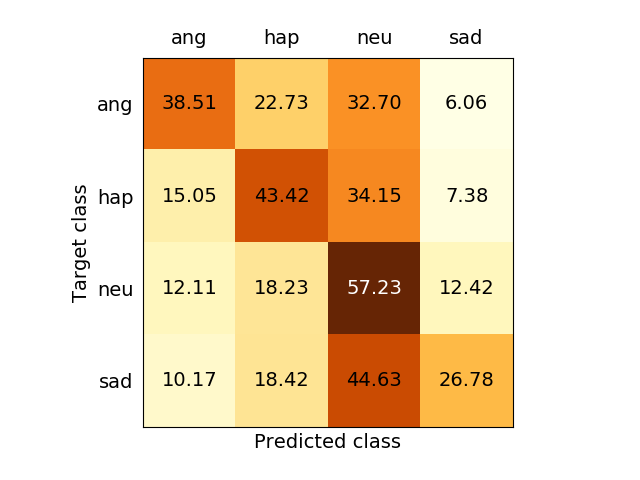}}
    \subfloat[FBank: 0dB SNR\label{fig:fbank_snr0}]{\includegraphics[trim=70 20 85 20, clip, width=.23\textwidth]{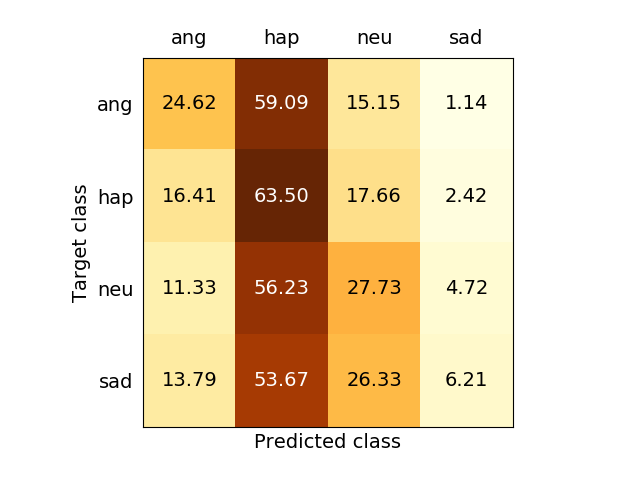}}
    \hfill
    \subfloat[FBank+Video: clean audio\label{fig:fbank_clean_multi}]{\includegraphics[trim=50 20 85 20, clip, width=.245\textwidth]{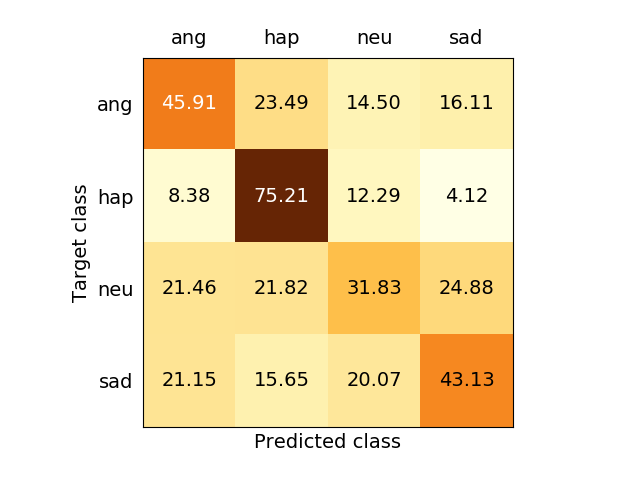}}
    \subfloat[FBank+Video: 0dB SNR\label{fig:fbank_snr0_multi}]{\includegraphics[trim=70 20 85 20, clip, width=.23\textwidth]{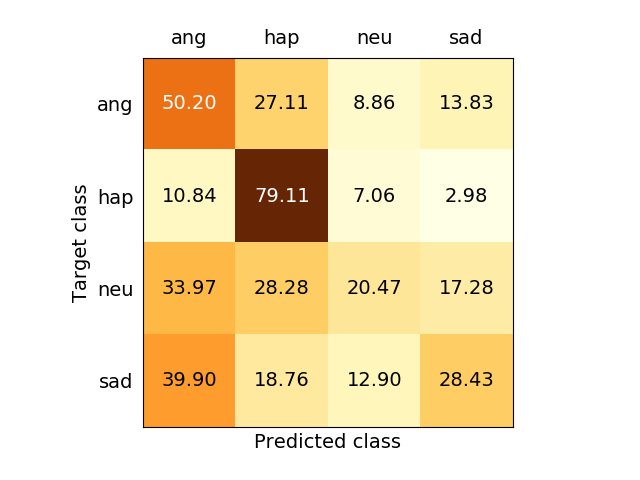}}

    \caption{Results (recall in \%) for MSP-IMPROV from uni- and multimodal models trained on clean audio only and tested on clean and noisy (0dB SNR) audio. (a) \& (b): audio-only with eGeMAPS features, (c) \& (d): audiovisual with eGeMAPS; (e)~\&~(f): audio-only with filterbank features, (g) \& (h): audiovisual with filterbanks.}
    \label{fig:confusion}
\end{figure*}

To gain more insights on the results, we analyzed the models' predictions. Exemplary confusion matrices are shown in Fig.~\ref{fig:confusion}.
For clean audio, the individual class recalls with eGeMAPS and DeepSpectrum features are well balanced (cf.~Fig.~\ref{fig:gemaps_clean}). For filterbanks we observed a bias towards the class \textit{neutral} (Fig.~\ref{fig:fbank_clean}). Adding visual features improves the recall for \textit{angry} and \textit{happy} notably, which can partially be explained by the high recall for \textit{happy} in the video-only case.
When applied to noisy audio, we observed that \textit{happy} is predominantly predicted for all three feature types. With eGeMAPS (audio-only), this effect is most pronounced (cf. Fig.~\ref{fig:gemaps_snr0}). Adding the visual modality improves recall for the other three classes significantly (cf. Fig.~\ref{fig:gemaps_snr0_multi}).
With filterbanks and DeepSpectrum features (audio-only), the majority of samples is predicted as either \textit{happy} or \textit{neutral} at low noise levels and the bias towards \textit{happy} increases with higher noise levels. Adding visual features improves the recall for \textit{angry} and \textit{sad} considerably (cf. Fig.~\ref{fig:fbank_clean}-\ref{fig:fbank_snr0_multi}), but a bias towards the high-arousal classes \textit{happy} and \textit{angry} remains at higher noise levels.

One main finding of the analysis is that the performance \textit{decline} on noisy data is smallest for the model with filterbank features. However, the reference performance on clean data is lowest in this case.
The confusion matrices show that the model with filterbank features is more stable on noisy audio data with respect to the balance between classes and the bias towards one single class. Fig.~\ref{fig:gemaps_snr0} and~\ref{fig:fbank_snr0} illustrate this comparison: While with GeMAPS features, the number of samples wrongly predicted as \textit{happy} is very high and almost all \textit{angry} samples are predicted as \textit{happy}, these effects are less pronounced with filterbank features. This difference between the models and features becomes even larger at high noise levels.

For \textbf{CREMA-D} (Fig.~\ref{fig:crema_clean}), the results show the same patterns as for MSP-IMPROV: performance decline proportional to the noise level (even more pronounced at higher noise levels). An exception is the combination of visual and eGeMAPS features, for which the UAR at 5dB SNR is higher than on clean audio. We found that the benefit of adding visual features is differently distributed across emotion classes. In general, the largest improvement is observed for the class \textit{happy}, which also appears to be the class with the highest recall for video-only. As a result, it can happen that the total UAR is slightly higher on noisy data than on clean data because the imbalance between classes increases and certain biases are even more emphasized.
This effect is not as strong for MSP-IMPROV because overall the differences in recall for individual classes are not as large.

The best audio-only result is 63.76\%~$\pm$1.35\% (clean audio, eGeMAPS), the best multimodal result is obtained with eGeMAPS features (71.38\%~$\pm$0.28\% at 5dB SNR and 71.17\%~$\pm$0.29\% on clean audio). The video-only UAR is 59.63\%~$\pm$0.71\%.

The inspection of confusion matrices showed high recall for the class \textit{angry} on clean audio throughout all feature sets. With eGeMAPS the class recalls are most balanced, while with filterbanks a high proportion of samples is wrongly predicted as \textit{neutral}.
On noisy audio we observed a strong bias towards \textit{happy} with eGeMAPS, a strong bias towards \textit{angry} with filterbanks and high confusion between \textit{sad} and \textit{neutral} with DeepSpectrum features. With higher noise levels the biases towards the high-arousal classes \textit{angry} and \textit{happy} become stronger.
The addition of visual features improves generally the recall for \textit{happy} and \textit{neutral}.




\subsection{Models trained on single noise levels}
\begin{figure}[ht]
    \centering
    \includegraphics[trim=10 10 10 10, clip, width=\linewidth]{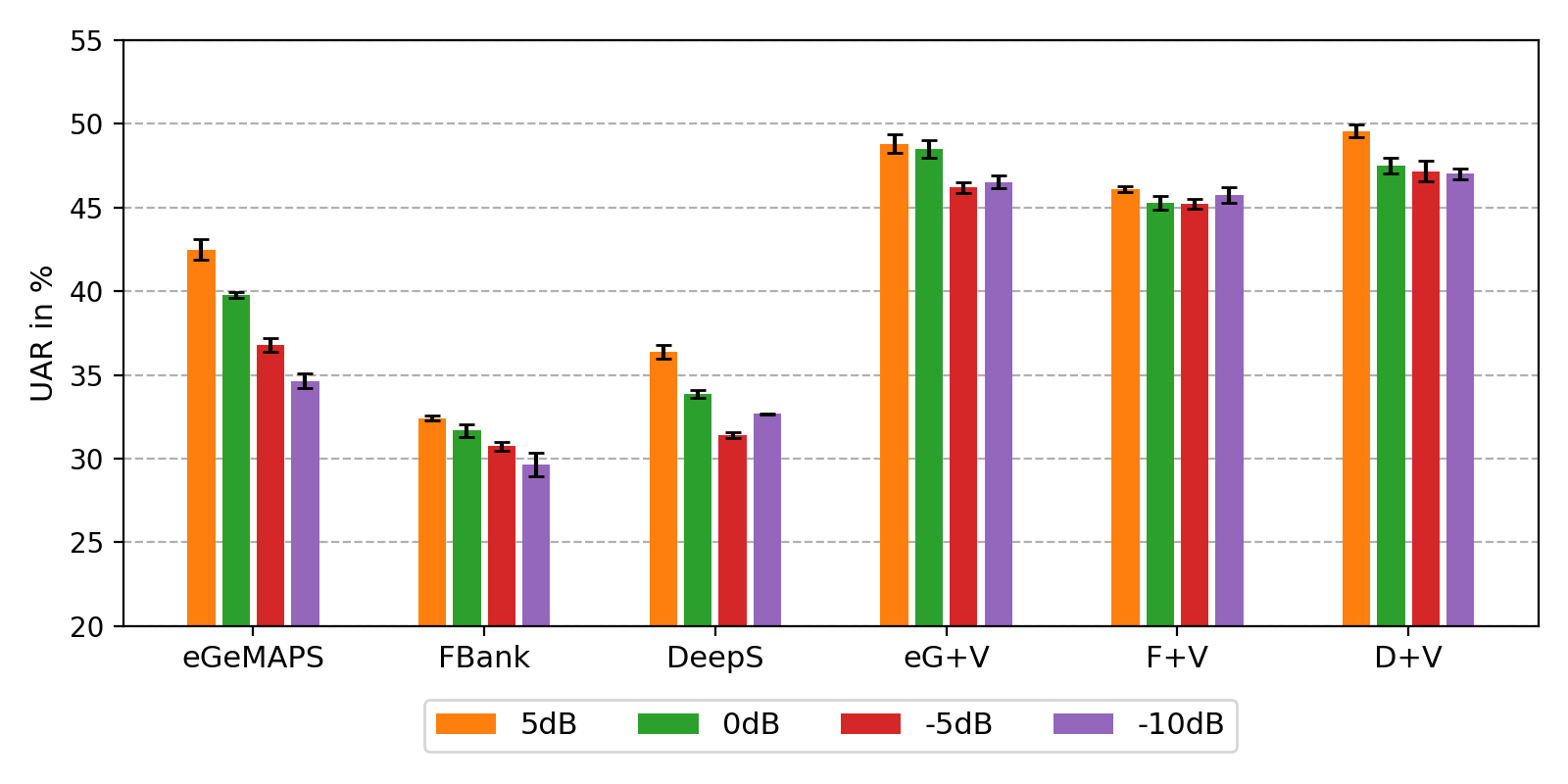}
    \caption{Results on MSP-IMPROV for training and evaluation on single noise levels (babble noise).}
    \label{fig:msp_each}
\end{figure}

In the second experiment, we train and evaluate the models at the same noise level (\textit{matched} condition). 
Fig.~\ref{fig:msp_each} shows the results for MSP-IMPROV. The results for CREMA-D exhibit similar characteristics.
In contrast to the first experiment the performance remains much more stable for noisy data when the model is trained on this kind of data. The results within one feature set are at a similar level with a tendency of slightly lower UAR for higher noise levels. This decrease is most pronounced for eGeMAPS features (audio-only).
These observations paired with the findings of the first experiment emphasize the severe consequences that a mismatch between train and test data can cause.

\subsection{Data augmentation: Training on all noise levels}
\begin{figure}[ht]
    \centering
    \includegraphics[trim=10 10 10 10, clip, width=\linewidth]{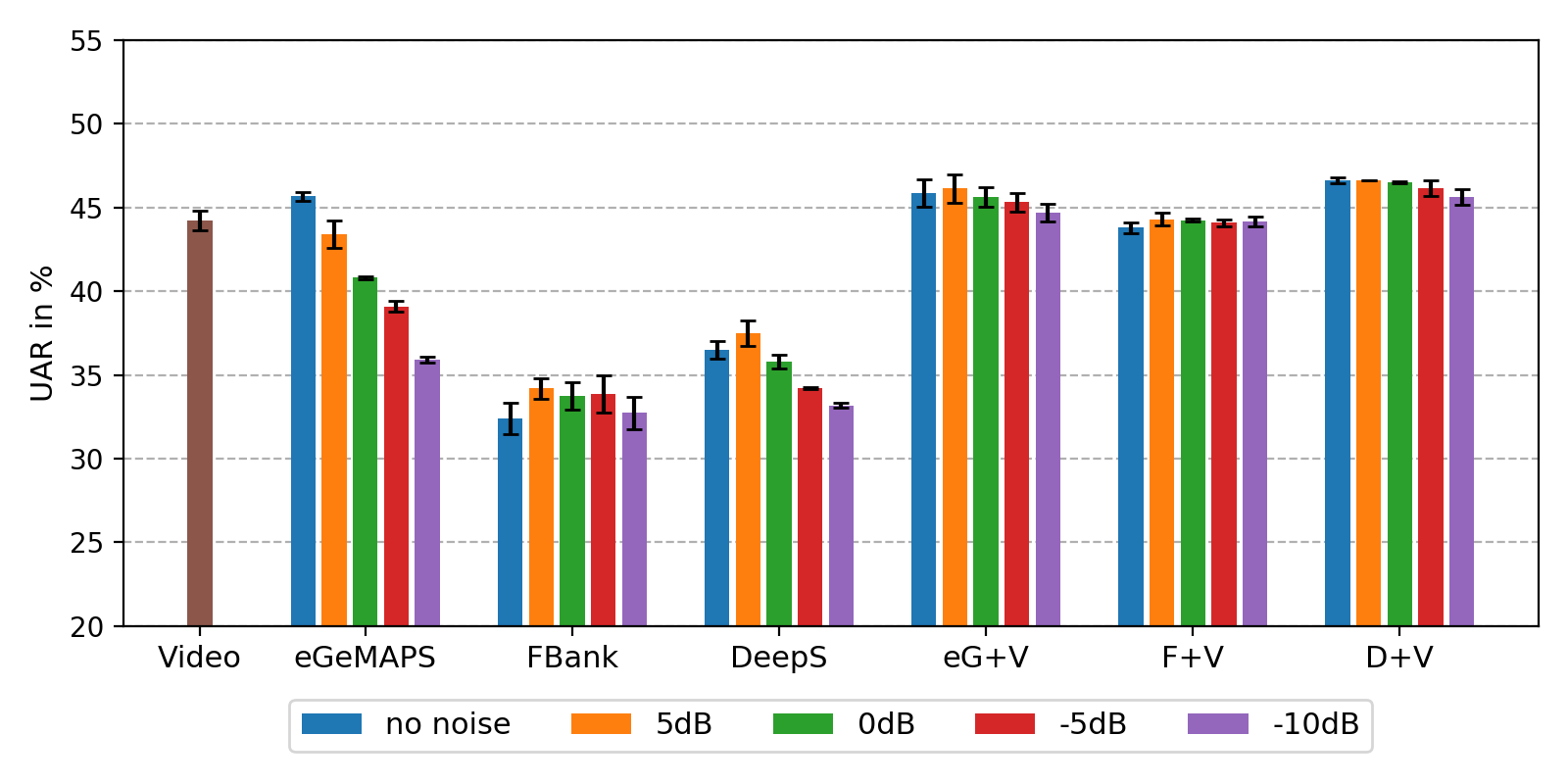}
    \caption{Results on MSP-IMPROV for training on all noise levels with babble noise (data augmentation).}
    \label{fig:msp_all}
\end{figure}

In the third experiment all models are trained on the union of data from all noise levels (including clean audio) and evaluated on the different noise levels separately (similar setup as in section~\ref{sec:exp1}).
The results for MSP-IMPROV are shown in Fig.~\ref{fig:msp_all}. Again, CREMA-D results exhibit similar characteristics.
Compared to Fig.~\ref{fig:msp_clean}, the performance on clean audio decreases throughout all features and multimodal combinations, while it improves remarkably on noisy data, especially for audio-only. The addition of visual features is especially useful for filterbank and DeepSpectrum features.
These results show that data augmentation in the form of added noise is beneficial in noisy conditions. However, a trade-off between lower accuracy on clean and higher accuracy on noisy data needs to be accepted.



\section{Conclusions}

In this investigation on audiovisual emotion recognition in noisy acoustic conditions, we have shown that the performance decreases significantly when training and test data do not match (clean vs. noisy), and that this effect is dampened with audiovisual models.
We showed that data augmentation by adding noise to the training set increases the accuracy on noisy audio significantly, but can affect  results on clean data negatively.

One limitation of the present study is that the Lombard effect -- the phenomenon that people speak differently than usual in noisy environments -- is not taken into account. To consider this, future work needs to be based on real-world noisy speech data instead of superimposed noise.

The comparison between feature sets showed that the eGeMAPS parameter set yields the best overall results.
However, the inspection of error patterns revealed that the CNN with log Mel filterbank features yields more stable predictions under noisy conditions with respect to the magnitude of accuracy decline and the class balance in predictions.




\bibliographystyle{IEEEbib}
\bibliography{bibliography}

\end{document}